\documentclass[reprint,twocolumn,aps,pra,amsmath,amssymb,superscriptaddress,nofootinbib]{revtex4-2}

\usepackage{float}
\usepackage{layouts}
\usepackage{graphicx}
\usepackage{color}
\usepackage{mathtools}
\usepackage{algpseudocode,algorithm,algorithmicx}

\begin{document}

\title{Demonstration of Optimal Non-Projective Measurement of Binary Coherent States with Photon Counting}
		
\author{M. T. DiMario}
\affiliation{Center for Quantum Information and Control, Department of Physics and Astronomy, University of New Mexico, Albuquerque, New Mexico 87131}
\affiliation{Joint Quantum Institute, National Institute of Standards and Technology and the University of Maryland, College Park, Maryland 20742}

\author{F. E. Becerra}
\email{fbecerra@unm.edu}
\affiliation{Center for Quantum Information and Control, Department of Physics and Astronomy, University of New Mexico, Albuquerque, New Mexico 87131}

\begin{abstract}

Quantum state discrimination is a central problem in quantum measurement theory, with applications spanning from quantum communication to computation. Typical measurement paradigms for state discrimination involve minimum probability of error or unambiguous discrimination with minimum probability of inconclusive results. Alternatively, an optimal inconclusive measurement, a non-projective measurement, achieves minimal error for a given inconclusive probability. This more general measurement encompasses the standard measurement paradigms for state discrimination and provides a much more powerful tool for quantum information and communication. Here, we experimentally demonstrate the optimal inconclusive measurement for the discrimination of binary coherent states using linear optics and \textcolor{black}{single photon detection}. Our demonstration uses coherent displacement operations based on interference, \textcolor{black}{single photon detection}, and fast feedback to prepare the optimal feedback policy for the optimal non-projective quantum measurement with high fidelity. This generalized measurement allows us to transition among standard measurement paradigms in an optimal way from minimum error to unambiguous measurements for binary coherent states. As a particular case, we use this general measurement to implement the optimal minimum error measurement for phase coherent states, which is the optimal modulation for communications under the average power constraint. Moreover, we propose a hybrid measurement that leverages the binary optimal inconclusive measurement in conjunction with sequential, unambiguous state elimination to realize higher dimensional inconclusive measurements of coherent states.
\end{abstract}

\maketitle

\section{Introduction}

Quantum measurement theory provides a fundamental understanding of the limits on the achievable sensitivity for distinguishing quantum states \cite{chefles00,barnett09,herzog04}. Physically realizable strategies that attain, or even approach, the ultimate sensitivity limits for distinguishing nonorthogonal coherent states have a wide range of applications in optical communication \cite{giovannetti04,loock08,guha11, rosati16,klimek16,banaszek20}, cryptography \cite{bennet84,bennett92,huttner95,grosshans02, silberhorn02,grosshans03,gisin02,sych10}, and quantum information processing \cite{munro05,nemoto04,ralph03}.
A central problem in quantum measurement theory and quantum information processing is the discrimination between two quantum states $|\psi_{1} \rangle$ and $| \psi_{2} \rangle$ with a certain optimal measurement given an optimality criterion, depending on the specific application \cite{barnett09,bergou04,Bae15}.

Two fundamental measurement paradigms for quantum state discrimination involve either minimum error or unambiguous state discrimination. Minimum-error state discrimination (MESD) aims to achieve minimal probability of error $P_{\mathrm{E}}$ \cite{yuen75,helstrom76,ban97,barnett97,tsujino11,wittmann08,izumi20b,becerra13,dimario18,becerra15,burenkov21,sidhu21}. The Helstrom bound \cite{helstrom76} gives the ultimate limit for $P_{\mathrm{E}}$, which is achieved by projective measurements onto complex superpositions of quantum states. Notably, the optimal MESD measurement for binary coherent states can be realized with linear optics, \textcolor{black}{single photon detection}, and fast feedback \cite{dolinar,cook07}. In contrast, unambiguous state discrimination (USD) allows for perfect discrimination with $P_{\mathrm{E}}=0$, but requires a non-zero probability of inconclusive results $P_{\mathrm{I}} \neq 0$. Such a non-projective measurement is described by a positive operator-valued measure (POVM) with three elements \cite{eldar03,barnett09,raynal03}, and aims to achieve the smallest possible $P_{\mathrm{I}}$  \cite{ivanovic87,dieks88,peres88,jaeger95,huttner96,peres98,becerra13b,izumi21,sidhu21b}. The realization of optimal USD of binary coherent states does not require feedback \cite{banaszek99,enk02,huttner95}, allowing for simpler implementations \cite{bartuskova08,becerra13b} compared to optimal MESD.

While optimal projective measurements exist for certain binary discrimination tasks \cite{helstrom76,barnett09,izumi18, takeoka06}, quantum measurement theory allows for a broader class of generalized quantum measurements which are not projective. These generalized measurements provide a more powerful tool for quantum information processing and communications \cite{barnett09}. Among these general quantum measurements, the optimal inconclusive measurement achieves the smallest possible error probability for a fixed probability of inconclusive results \cite{chefles98b, eldar03}. This measurement is a non-projective measurement, and thus described by a non-projective POVM, that encompasses MESD and USD measurement paradigms. Moreover, non-projective quantum measurements allow for more exotic discrimination tasks such as quantum state elimination \cite{crickmore20}, state comparison \cite{andersson06, sedlak07,barnett03}, and discrimination with a fixed error margin \cite{sugimoto09}. Furthermore, understanding optimal inconclusive measurements for binary states may provide a path for realizing arbitrary non-projective POVMs in a two dimensional Hilbert space \cite{nakahira18, izumi20}.
\begin{figure*}[t]
	\includegraphics[width = \textwidth]{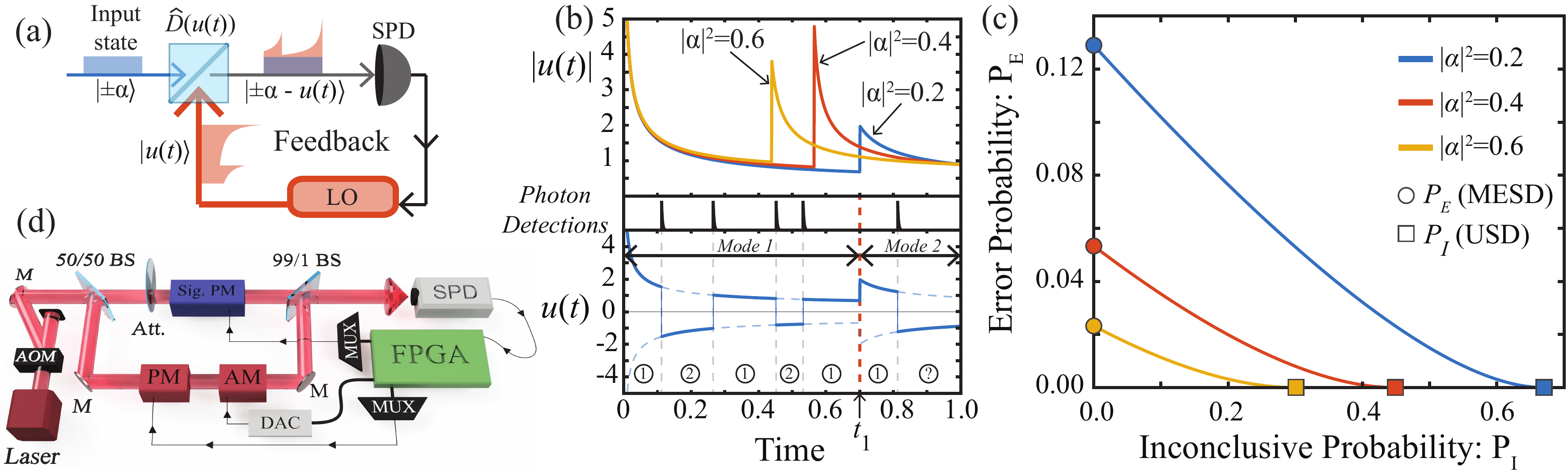}
	\caption{(a) Schematic of the generalized receiver for the optimal inconclusive measurement. The input states are displaced in phase space using an optimal waveform $u(t)$ for the LO field and followed by a single photon detector (SPD) and feedback operations. (b) Optimal waveform magnitudes $|u(t)|$ for different mean photon numbers (top panel). The bottom panel shows an example of the waveform $u(t)$ for a particular measurement record where the phase switches between $0$ and $\pi$ every photon detection (\textcolor{black}{Photon Detections}, middle panel). The circles along the $x-$axis show the current hypothesis for the input state as the measurement progresses. (c) Error probability $P_{\mathrm{E}}$ for the optimal inconclusive measurement as a function of specified probability of inconclusive results $P_{\mathrm{I}}$ for $|\alpha|^{2}$=0.2, 0.4, and 0.6. \textcolor{black}{The colored circles along the $y$-axis ($P_{\mathrm{I}}=0$) correspond to smallest possible $P_{\mathrm{E}}$ in the paradigm of MESD, and the colored squares along the $x$-axis ($P_{\mathrm{E}}=0$) correspond to the smallest possible $P_{\mathrm{I}}$ in the paradigm of USD.} (d) Experimental setup used for demonstrating the optimal inconclusive measurements of binary coherent states.}
	\label{optincth}
\end{figure*}

Theoretical work on quantum measurement theory has shown that it is possible to realize an optimal inconclusive measurement for a broad class of quantum states based on local operations and classical communications \cite{sugimoto09,fiurasek03, hayashi08}. However, the corresponding measurement operators for the discrimination of optical coherent states do not necessarily have a feasible physical realization. While suboptimal inconclusive measurements of coherent states can be realized based on linear optics and \textcolor{black}{single photon detection} \cite{wittmann10}, their performance falls short of the performance of the optimal inconclusive measurement. Recent work in Ref. \cite{nakahira12} proposed a physical realization of a strategy for the optimal inconclusive measurement of binary coherent states. It was shown that such a non-projective measurement can be realized using displacement operations, \textcolor{black}{single photon detection}, and feedback \cite{nakahira12}, which are the same physical elements needed for implementing arbitrary binary projective measurements \cite{takeoka05, takeoka06}.

In this work, we experimentally demonstrate the optimal inconclusive measurement for binary coherent states \cite{nakahira12}. The measurement splits the energy of the input state into two temporal modes. It performs a MESD measurement in the first mode providing conclusive results with a certain probability of error, and an optimal inconclusive measurement in the single-state domain in the second mode determining whether the measurement result is inconclusive. Our demonstration uses low noise, high bandwidth real-time feedback conditioned on single photon detections to prepare the optimal displacement operations required for the optimal inconclusive measurement. We further use this generalized optimal measurement to realize the optimal MESD for phase coherent states, which is the optimal modulation for optical communications under the average power constraint, thus demonstrating the optimal quantum receiver for coherent optical communications. Lastly, we show that the binary optimal inconclusive measurement enables the realization of inconclusive discrimination of three coherent states when used together with measurements for unambiguous state elimination based on with hypothesis testing. This proposed method can in principle be extended to high dimensional inconclusive measurement strategies of coherent states.
\section{Results}
\subsection{Optimal Inconclusive Measurement}
The optimal inconclusive measurement is a non-projective quantum measurement that encompasses the MESD and USD paradigms and optimizes the tradeoff between errors and inconclusive results \cite{chefles98b,eldar03}. By construction, the optimal inconclusive measurement achieves the minimal error probability $P_{\mathrm{E}}$ for a specified inconclusive probability $P_{\mathrm{I}}$ \cite{nakahira12,barnett09}. A feasible realization of the POVM for the optimal inconclusive measurement $\{\hat{\Pi}_{1}, \hat{\Pi}_{2}, \hat{\Pi}_{?}\}$ for binary coherent states was recently proposed in \cite{nakahira12}. Notably, this optimal non-projective measurement can in principle be realized by a generalization of the optimal receiver for MESD, called the Dolinar receiver. This optimal MESD receiver is based on displacement operations in phase space implemented by interfering the input state with a local oscillator (LO) field, \textcolor{black}{single photon detection}, and feedback with an optimal feedback policy. The displacement has a magnitude with an optimal waveform and a phase conditioned to photon detection \cite{dolinar,cook07,geremia04b}.

Figure \ref{optincth}(a) shows the concept of the optimal inconclusive measurement. The input state $|\pm \alpha \rangle$ and strong LO field interfere on a high transmittance beam splitter to implement a displacement operation $\hat{D}(u(t))$. The receiver implements the optimal displacement waveform $u(t)$, where the phase of the LO switches between 0 and $\pi$ based on the photon detection outcomes from the single photon detector (SPD) during the measurement time. In the proposal for the optimal inconclusive discrimination strategy \cite{nakahira12}, the generalized receiver performs optimal measurements in two temporal modes during the measuring time $0 \leq t \leq 1$ using displacement operations, \textcolor{black}{single photon detection}, and feedback. In the first temporal mode ($0 \leq t \leq t_{1}$), the receiver performs an optimal MESD measurement to discriminate between $\{|\pm\alpha\rangle\}$ with minimal error using the optimal displacement waveform \cite{dolinar,cook07,geremia04b,nakahira12}. In the second temporal mode ($t_{1}<t\leq1$), the receiver performs an optimal inconclusive measurement in the so-called single-state domain, where the measurement becomes a projective measurement such that the POVM element for the least probable is zero, e.g. $\hat{\Pi}_{2}=0$ \cite{nakahira12,sugimoto09}. Without loss of generality, the most probable state after the first mode is $|\alpha\rangle$, and the non-zero POVM elements are $\hat{\Pi}_{1}$ and $\hat{\Pi}_{?}$. Therefore, the receiver in the second temporal mode attempts to determine whether the measurement result is inconclusive, i.e. the receiver realizes a MESD measurement between the correct and inconclusive outcomes. Reference \cite{nakahira12} shows that this projective measurement in the second temporal mode (single-state domain) can be realized by a Dolinar-\textit{like} receiver with a different optimal displacement waveform, which is the key element that we leverage for demonstrating the optimal inconclusive measurement. The total displacement waveform $u(t)$ that implements the optimal inconclusive measurement is given by \cite{nakahira12}:
\begin{equation}
	u(t) =
	\begin{cases}
		\frac{(-1)^{N_{1}(t)}\alpha}{\sqrt{1 - 4p(1-p)K^{2t}}} & 0\leq t \leq t_{1}\\
		\frac{(-1)^{N_{2}(t)+N_{0}}\alpha}{\sqrt{1 - 4v(1-v)K^{2(t - t_{1})}}} & t_{1} < t\leq 1.
	\end{cases}
	\label{totalwf}
\end{equation}
\noindent \textcolor{black}{Where $N_{1}(t)$ and $N_{2}(t)$ are the total number of photons detected up to time $t$ for the first and second temporal mode, respectively, and $N_{0} \in \{0,1\}$ based on $|\alpha|^{2}$, $P_{\mathrm{I}}$, and $p$ \cite{nakahira12}. The total optimal waveform $u(t)$ is comprised of $u_{1}(t)$ and $u_{2}(t)$, each of which are optimal for the two temporal modes (see Methods Section I for details). The magnitude of $u(t)$ is predetermined based on the values of $|\alpha|^{2}$, $P_{\mathrm{I}}$, and $p$, but the sign of $u(t)$ (phase of the LO) is adaptively switches between positive and negative (LO phase of 0 and $\pi$) each photon detection due to the $(-1)^{N_{1}(t)}$ and $(-1)^{N_{2}(t) + N_{0}}$ terms.}
\begin{figure*}[t]
	\includegraphics[width = \textwidth]{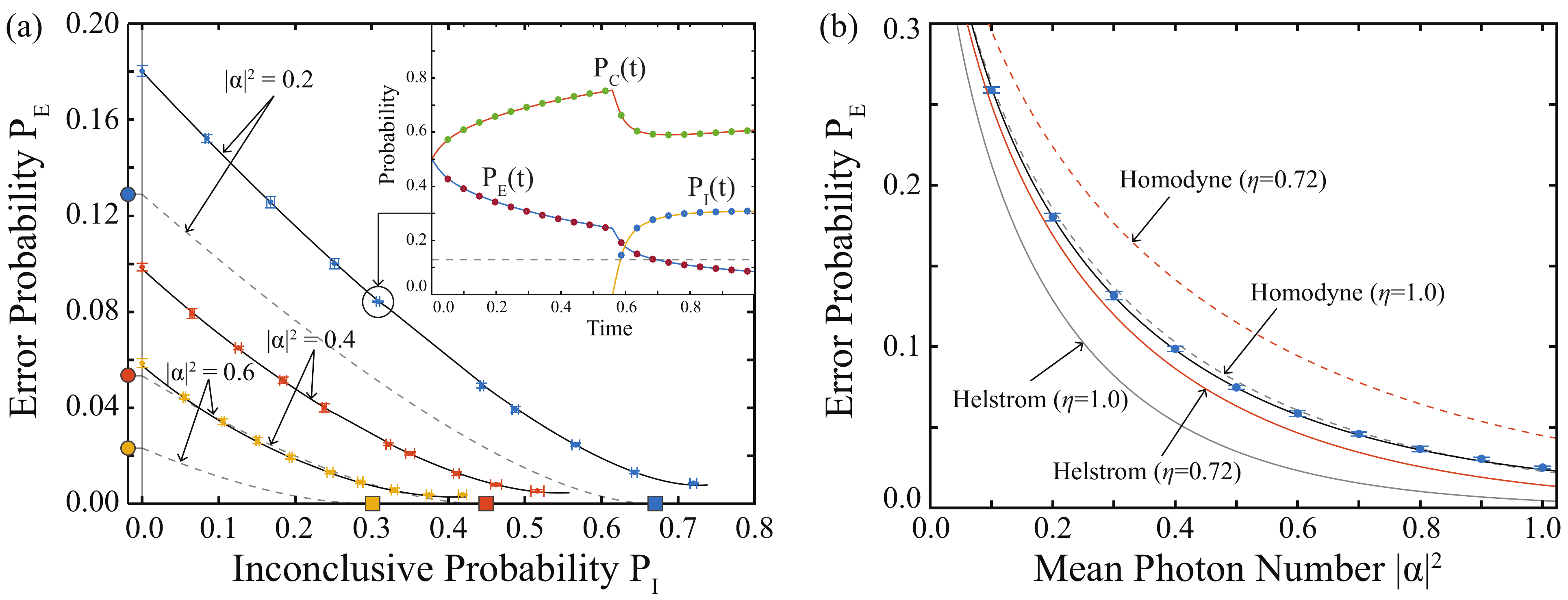}
	\caption{(a) Experimental results for the optimal inconclusive measurement for $|\alpha|^{2}$=0.2, 0.4, and 0.6, in blue, orange, and yellow, respectively. Each point corresponds to the measured values of $\{P^{\mathrm{exp}}_{\mathrm{I}}(1),P^{\mathrm{exp}}_{\mathrm{E}}(1)\}$ and the error bars represent one standard deviation from five sets of $5\times 10^{4}$ individual experiments each. The solid lines show the expected results and the dashed gray lines show the ideal performance for each $|\alpha|^{2}$. The colored circles and squares on the $y$-axis and $x$-axis show the optimal $P_{\mathrm{E}}$ and $P_{\mathrm{I}}$ for ideal MESD and USD, respectively. Inset (i): evolution of $P^{\mathrm{exp}}_{\mathrm{E}}$, $P^{\mathrm{exp}}_{\mathrm{C}}$, and $P^{\mathrm{exp}}_{\mathrm{I}}$ for $|\alpha|^{2}$=0.2 and and $P^{\mathrm{exp}}_{\mathrm{I}}$=0.31. (b) Experimental results (blue points) of an optimal MESD measurement, the Dolinar receiver, for phase coherent states $\{|\pm \alpha \rangle\}$. The gray and red solid lines show the Helstrom bound for $\eta=1.0$ and $\eta=0.72$, respectively, and the dashed lines show the corresponding error for a homodyne measurement.}
	\label{optincdol}
\end{figure*}

The top panel of Fig. \ref{optincth}(b) shows the displacement magnitude $|u(t)|$ for the optimal inconclusive strategy with inconclusive probability $P_{\mathrm{I}} = 0.19$ for $|\alpha|^{2}$ = 0.2, 0.4, and 0.6. The discrete jumps in $|u(t)|$ for each $|\alpha|^{2}$ correspond to the time $t_{1}$ when the receiver switches between the measurements two temporal modes. The receiver implements a minimum-error measurement with a Dolinar receiver during $0 \leq t \leq t_{1}$ with $u_{1}(t)$ between $|\pm \alpha \rangle$. The receiver then implements the optimal inconclusive measurement in the single-state domain $\{\hat{\Pi}_{1}, \hat{\Pi}_{?}\}$ with a Dolinar-\textit{like} receiver during $t_{1} < t \leq 1$ with $u_{2}(t)$. The final outcome of the measurement is either an inconclusive result with probability $P_{\mathrm{I}}$, a correct discrimination result with probability $P_{\mathrm{C}}$, or an error with probability $P_{\mathrm{E}} = 1 - P_{\mathrm{C}} - P_{\mathrm{I}}$ \cite{nakahira12}. The bottom panel of Fig. \ref{optincth}(b) shows the displacement amplitude $u(t)$ for an example measurement record. The provisional hypothesis (circles) and the phase of the waveform change each time a photon is detected. The red dashed line ($t_{1}\approx0.70$) shows where the receiver switches from MESD of the two input states in the first temporal mode, to MESD between the more likely state given the current detection record and the inconclusive outcome in the second temporal mode.

Figure \ref{optincth}(c) shows the resulting probabilities $\{P_{\mathrm{I}}, P_{\mathrm{E}}\}$ of the optimal inconclusive measurement for equiprobable coherent states $\{|\pm\alpha\rangle\}$ with $|\alpha|^{2}$ = 0.2, 0.4, and 0.6, in blue, orange, and yellow, respectively. \textcolor{black}{The colored circles along the $y$-axis ($P_{\mathrm{I}}=0$) correspond to the smallest possible $P_{\mathrm{E}}$ in the paradigm of MESD (Helstrom Bound), and the colored squares along the $x$-axis ($P_{\mathrm{E}}=0$) correspond to the smallest possible $P_{\mathrm{I}}$ in the paradigm of USD (sometimes referred to as the IDP Bound \cite{ivanovic87,dieks88,peres88}).} Thus, the optimal inconclusive measurement is the generalization of MESD and USD and interpolates between these measurement paradigms in an optimal way using more general non-projective measurements. In general, optimal inconclusive measurement for the discrimination of two general quantum states $\{| \psi_{1} \rangle,| \psi_{2} \rangle\}$ is represented by three POVM elements $\{\hat{\Pi}_{1}, \hat{\Pi}_{2}, \hat{\Pi}_{?}\}$ where a positive outcome of $\{\hat{\Pi}_{1,2}\}$ indicates that the state $| \psi_{1,2} \rangle$ is present, and $\hat{\Pi}_{?}=\hat{I}-\hat{\Pi}_{1}-\hat{\Pi}_{2}$ corresponds to an inconclusive result. Optimality indicates that this non-projective measurement achieves the minimum error for a fixed probability of inconclusive outcomes.

While the proposed implementation of this non-projective quantum measurement in Ref. \cite{nakahira12} is in principle feasible, its demonstration requires a high degree of control for the preparation of optimal waveforms with high fidelity, and the ability to realize feedback measurements with high bandwidth and low noise (see Supplementary Note I). Moreover, the validation of optimal performance requires absolute power measurements at the single photon level. In our experimental demonstration we address the issues to satisfy these stringent requirements, which allows us to demonstrate experimentally this complex quantum measurement with high fidelity. Figure \ref{optincth}(d) shows our experimental setup for the demonstration of the optimal inconclusive measurement for binary coherent states. We use an interferometric setup to generate the input states and local oscillator field, a single photon detector (SPD), and an FPGA \textcolor{black}{(Altera Cyclone IV, 50 MHz base clock)} connected to a digital-to-analog converter (DAC) to implement the required optimal displacement waveform $u(t)$ for the optimal inconclusive measurement using fiber-coupled amplitude (AM) and phase (PM) modulators (see Methods Sections B, C, and E for details). \textcolor{black}{We actively stabilize our interferometer using a second 780 nm laser and a feedback loop to maintain a well-defined relative phase (see Methods Section C for details). Our implementation achieves an overall detection efficiency $\eta=0.72(1)$ ($\eta=\eta_{\mathrm{SPD}}\eta_{\mathrm{sys}}$ where $\eta_{\mathrm{SPD}}=0.82(1)$ is the SPD efficiency and $\eta_{\mathrm{sys}}=0.88(1)$ is the system transmittance), interference visibility $\xi=0.998(1)$, and dark counts $\nu=0.03(1)$ per pulse. The experiment operates at a 4 kHz repetition rate, alternating between experimental trials (1024 time bins, 160 ns each) and interferometer stabilization with a $\approx 66 \%$ duty cycle. We have also realized numerical investigations of the effects of realistic imperfections described in Supplementary Note I. Based on these studies, we observe that reduced detection efficiency degrades the achievable performance for all error and inconclusive probabilities. Reduced interference visibility and increased dark counts mainly degrades the performance of strategies where the desired $P_{\mathrm{E}}$ or $P_{\mathrm{I}}$ are small, i.e. near the MESD and USD regimes.}
\subsection{Demonstration of the Optimal Inconclusive Measurement}
We implement the optimal inconclusive measurement for equiprobable coherent states. In our experimental demonstration, we obtain the time evolution of the error $P^{\mathrm{exp}}_{\mathrm{E}}(t)$, correct $P^{\mathrm{exp}}_{\mathrm{C}}(t)$, and inconclusive $P^{\mathrm{exp}}_{\mathrm{I}}(t)$ probabilities by reconstructing the results in post-processing, and compare them to the expected probabilities $\{P_{\mathrm{E}}(t), P_{\mathrm{C}}(t), P_{\mathrm{I}}(t)\}$ \cite{nakahira12}. The final inconclusive $P^{\mathrm{exp}}_{\mathrm{I}}(t=1)$ and error $P^{\mathrm{exp}}_{\mathrm{E}}(t=1)$ probabilities for a given $|\alpha|^{2}$ correspond to a single realization of the optimal inconclusive measurement. Figure \ref{optincdol}(a) shows the experimental results for $|\alpha|^{2}$ = 0.2, 0.4, and 0.6 in blue, orange, and yellow, respectively. The points show the experimental data $\{P^{\mathrm{exp}}_{\mathrm{I}}(1), P^{\mathrm{exp}}_{\mathrm{E}}(1)\}$ and the error bars represent one standard deviation from five experimental runs of $5\times10^{4}$ independent experiments each. The black lines show the theoretical expectation from Monte-Carlo simulations of the experiment incorporating experimental imperfections (See Supplementary Note IV). We note that we obtain the expected performance of our demonstration by directly simulating the experiment including experimental imperfections and effects without requiring ant fitting procedures. The dashed gray lines show the ideal ($\eta=1$) performance for each mean photon number. The colored circles and squares on the $y$-axis and $x$-axis show the optimal $P_{\mathrm{E}}$ and $P_{\mathrm{I}}$ for ideal MESD and USD, respectively, for each $|\alpha|^{2}$.

We observe that our demonstration of the optimal inconclusive measurement with $\eta=0.72$ for $|\alpha|^{2}$ = 0.2, 0.4, and 0.6 reaches errors below the ideal Helstrom bound when $P_{\mathrm{I}} \gtrapprox 0.18$. This shows that a non-ideal implementation of the optimal inconclusive measurement can surpass the ideal Helstrom bound at the expense of having inconclusive results\footnote{\textcolor{black}{We note that while the Helstrom bound is the minimum discrimination error that can be achieved by a deterministic measurement, this bound is not the lowest error for a general quantum measurement that allows for inconclusive results \cite{chefles98b}. As such, the optimal inconclusive measurement then allows for errors below the Helstrom bound for $P_{\mathrm{I}}\neq 0$, and achieves zero error at a rate of inconclusive results given by the IDP bound \cite{ivanovic87,dieks88,peres88}.}}. The inset of Figure \ref{optincdol}(a) shows an example of the evolution of $P_{\mathrm{E}}(t)$ (blue), $P_{\mathrm{C}}(t)$ (orange), and $P_{\mathrm{I}}(t)$ (yellow) as the measurement progresses for $|\alpha|^{2}=0.2$ and $P_{\mathrm{I}}\approx0.31$. The solid lines show the theoretical expectation including experimental imperfections and the points show the experimental results for $P^{\mathrm{exp}}_{\mathrm{E}}(t)$, $P^{\mathrm{exp}}_{\mathrm{C}}(t)$, and $P^{\mathrm{exp}}_{\mathrm{I}}(t)$ every 50 time bin steps. Note that the measurement switches from a MESD measurement to an optimal inconclusive measurement in the single-state domain at  $t_{1}\approx0.57$.  
\subsection{Optimal MESD of Binary Phase States}
The optimal inconclusive measurement generalizes MESD and USD \cite{nakahira12}, and can be used to demonstrate the optimal MESD measurement, the Dolinar receiver  \cite{dolinar}, by setting $P_{\mathrm{I}}=0$. Previous work \cite{cook07} demonstrated a Dolinar receiver for intensity-modulated coherent states $\{|0\rangle,|\alpha\rangle\}$ and achieved performance below the shot noise after correcting for system losses and  detection efficiency. However, phase encoded coherent states $\{|\pm\alpha \rangle\}$ are the optimal modulation for binary coherent communications under the energy constraint. This is because this alphabet has the smallest overlap, and therefore the highest distinguishability, for a fixed average energy of the states \cite{takeoka08, weedbrook12, banaszek99}. To this end, we use the optimal inconclusive measurement to demonstrate a Dolinar receiver for the phase encoded binary coherent states,  and thus demonstrate the optimal quantum receiver for coherent optical communications.

Figure \ref{optincdol}(b) shows the experimental results (blue points) and the expected error probability (solid black) for the optimal MESD measurement, the Dolinar receiver, for phase coherent states together with the Helstrom (solid) and homodyne limits (dashed). The Helstrom bound and the homodyne limits corrected to our overall efficiency ($\eta=0.72$) are included for reference. We observe that our demonstration of the Dolinar receiver approaches the corrected Helstrom bound and shows an excellent agreement with the theoretical predictions (solid black line). \textcolor{black}{We note that our implementation of the optimal MESD measurement for BPSK states with overall efficiency of $\eta=0.72$ achieves $P_{\mathrm{E}}=0.18$ for $|\alpha|^{2}=0.2$, which is below the ideal homodyne limit that corresponds to the optimal Gaussian measurement for BPSK \cite{takeoka08}. This error rate is similar to the one achieved by a sub-optimal receiver without feedback in Ref. \cite{tsujino11} using a high efficiency ($\eta=0.99$) superconducting detector resulting on an overall system efficiency of $\eta=0.91$. Thus, we conclude the strategy demonstrated here based on complex adaptive measurements can potentially provide overall higher sensitivities than the sub-optimal strategy under the same loss and realistic experimental noise and imperfections.} In principle the optimal inconclusive measurement also allows for construction of the optimal USD measurement where $P_{\mathrm{E}}=0$. However, the experimental imperfections such as dark counts and non-ideal interference visibility prevent the receiver from achieving $P_{\mathrm{E}}=0$, see Fig. \ref{optincdol}(a). Nevertheless, the above framework allows for finding the optimal waveform to implement this optimal measurement.
\subsection{Higher Dimensional Inconclusive Strategies}
\begin{figure}[b]
	\includegraphics[width = 8.5cm]{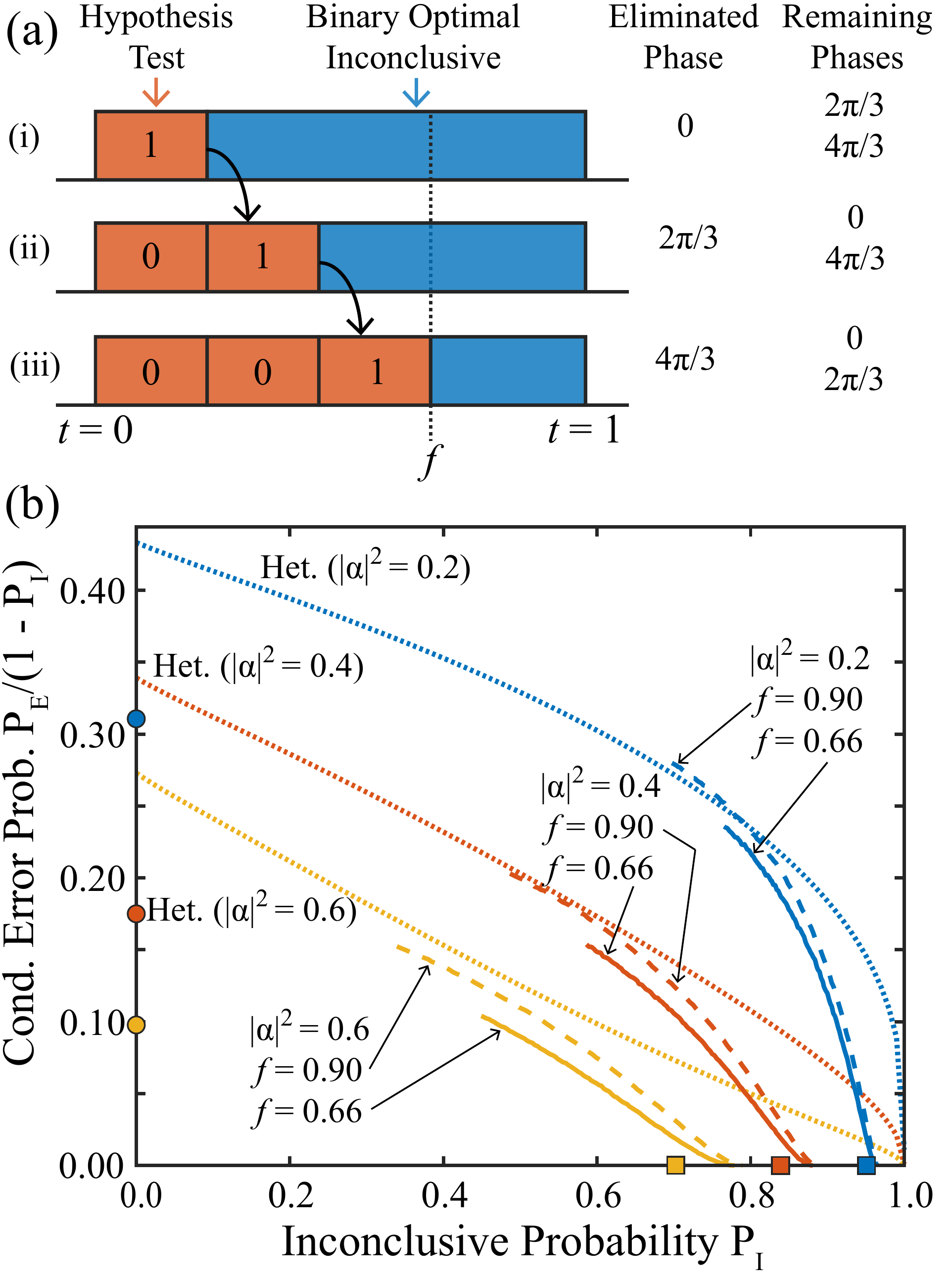}
	\caption{Inconclusive discrimination of the ternary phase-shift keyed (TPSK) states coherent states using the binary optimal inconclusive measurement. (a) The proposed inconclusive measurement for TPSK states uses sequential unambiguous state elimination, followed by the optimal binary inconclusive measurement (See main text for details). (b) Conditional error probability $P_{\mathrm{E}}/(1 - P_{\mathrm{I}})$ as a function of inconclusive probability $P_{\mathrm{I}}$ for $|\alpha|^{2} = $ 0.2 (blue), 0.4 (orange), and 0.6 (yellow). We compare the proposed measurement with the parameter $f$ = 0.66 (solid) and $f$ = 0.90 (dashed) to the performance of ideal heterodyne detection (dotted). See text for details.}
	\label{tpsk_oim}
\end{figure}
We investigate how to leverage the optimal inconclusive measurement of binary coherent states to enable inconclusive state discrimination of higher dimensional encodings. We propose a hybrid measurement that utilizes binary optimal inconclusive measurements in conjunction with unambiguous state elimination, which can realize such a non-projective  inconclusive measurement of ternary phase-shift keyed (TPSK) $\{ |\alpha \rangle, |\alpha e^{i2\pi/3} \rangle, |\alpha e^{i4\pi/3} \rangle \}$, and can be extended to higher dimensions.  This measurement first aims to eliminate all but two possible input states via hypothesis testing, and then utilize the optimal inconclusive measurement in the reminding binary states. Figure \ref{tpsk_oim}(a) shows the measurement operations conditioned to \textcolor{black}{single photon detection}, realized by the proposed high-dimensional inconclusive measurement. The receiver realizes an elimination measurement based on hypothesis testing (red region, Fig. \ref{tpsk_oim}(a-$i$)) for the state $|\alpha\rangle$ on a fraction $f/3$ of the total input state. This state elimination measurement is based on displacement operation of $|\alpha\rangle$ to the vacuum state $|0\rangle$ and \textcolor{black}{single photon detection}, such that detection of a photon unambiguously eliminates $|\alpha\rangle$ as a possible input state. If a photon is detected in the first stage (Stage 1), the receiver then performs an optimal inconclusive measurement (blue region, Fig. \ref{tpsk_oim}(a-$i$)) to optimally discriminate between $|\alpha e^{i2\pi/3}\rangle$ and $|\alpha e^{i4\pi/3}\rangle$ using the remaining fraction $1-f/3$ of the input energy. If no photons are detected during the first stage, the receiver then realizes a state elimination measurement for the input state $|\alpha e^{i2\pi/3} \rangle$ also using a fraction $f/3$ of the total input power (Fig. \ref{tpsk_oim}(a-$ii$)) . Now if a photon is detected in the second stage (Stage 2), an optimal inconclusive measurement discriminates between the remaining two possible input states using a fraction $1-2f/3$ of the input power, where the factor of 2 comes from the first stage. If no photons are detected in the second stage, then the receiver tests for the state $|\alpha e^{i4\pi/3} \rangle$ in Stage 3, also using a fraction $f/3$ of the total input state (Fig. \ref{tpsk_oim}(a-$iii$)) . If a photon is detected, an optimal inconclusive measurement discriminates between the remaining two states now using a fraction $1-3f/3$ of the input power. If no photons are detected in the third hypothesis test for unambiguous state elimination, we define the measurement outcome to be inconclusive.

Figure \ref{tpsk_oim}(b) shows the simulation results for the proposed inconclusive measurement of TPSK states based on the optimal inconclusive measurement for binary states for mean photon numbers $|\alpha|^{2} =$ 0.2, 0.4, and 0.6. The $x$-axis corresponds to the inconclusive probability and the $y$-axis corresponds to the conditional error probability $P_{\mathrm{E}}/(1 - P_{\mathrm{I}})$, i.e. the error probability given that a conclusive outcome was obtained. The solid and dashed lines show the proposed inconclusive measurement of three coherent states with $f=0.66$ and $f=0.90$, respectively. The dotted lines show the result $P^{\mathrm{Het}}_{\mathrm{E}}/(1 - P^{\mathrm{Het}}_{\mathrm{I}})$ for using ideal heterodyne detection, where measurement outcomes with the largest error probability are designated as inconclusive until the desired inconclusive probability is achieved as in \cite{becerra13b}.  \textcolor{black}{The optimal limits for USD and MESD for three states are represented with squares and circles in the \textit{x}-axis ($P_{\mathrm{E}}=0$) and \textit{y}-axis ($P_{\mathrm{I}} = 0$), respectively \cite{dallapozza15,enk02}.}

\textcolor{black}{The overall $P_{\mathrm{I}}$ achieved by this strategy contains two contributions $P_{\mathrm{I}} = P^{(1)}_{\mathrm{I}} + P^{(2)}_{\mathrm{I}}$, where $ P^{(1)}_{\mathrm{I}}$ comes from the state elimination stage, and $ P^{(2)}_{\mathrm{I}}$ from the binary optimal inconclusive measurement. In the state elimination stage, the detection of vacuum during all three hypothesis tests results in an inconclusive outcome as each state is equally likely. This produces a lower bound ($ P^{(1)}_{\mathrm{I}}$) on the attainable inconclusive probability $P_{\mathrm{I}}$ depending on the values of $f$ and $|\alpha|^{2}$. In the binary optimal inconclusive stage, the proposed measurement defines a “target” inconclusive probability ($ P^{(2)}_{\mathrm{I}}$) which can be set for any $|\alpha|^{2}$ to achieve the desired overall $P_{\mathrm{I}}$ of the strategy.} We observe that the proposed measurement with both values of the parameter $f$, which parameterizes the unambiguous state elimination stages, can outperform heterodyne detection.  Moreover, we note that a smaller value of $f=0.66$ achieves a smaller error probability, but this also puts a limit on the smallest attainable inconclusive probability of $P_{\mathrm{I}}\approx$0.76, 0.58, and 0.45 for $|\alpha|^{2}=$ 0.2, 0.4, and 0.6, respectively. On the other hand, a larger value of $f=0.90$ (dashed lines) allows for a smaller inconclusive probability but at the cost of a larger error probability compared to $f=0.66$ (solid lines). This trade-off is due to the fact that a larger value of $f$ results in a smaller contribution to $P_{\mathrm{I}}$ from the inconclusive outcome during the state elimination stage (detecting vacuum during all stages). However a larger value of $f$ results in a smaller fraction of the total input energy of the state for the binary optimal inconclusive measurement, which results in a larger error probability. \textcolor{black}{Then, the optimal choice of energy fraction $f$ will depend on the particular application of this proposed measurement. For example, if we are willing to tolerate more inconclusive results $P_{\mathrm{I}}$ to achieve a given small target error threshold $P_{\mathrm{E}}$ such as for communications with error detection, correction, and erasures \cite{chen12}, we should choose a small value of $f$.}

The proposed inconclusive measurement for three states can be extended to higher dimensions. Using this technique, an inconclusive measurement of $M$ input coherent states can be realized through the implementation of $M-1$ hypothesis testing stages for unambiguous state elimination \cite{enk02, becerra13b}, followed by the binary optimal inconclusive measurement. Given that the binary optimal inconclusive measurement can always achieve $P_{\mathrm{E}}=0$ for $P^{(2)}_{\mathrm{I}}<1$, there is always a range of error probabilities for which this strategy will outperform heterodyne detection (note $P^{\mathrm{Het}}_{\mathrm{E}}=0$ only when $P^{\mathrm{Het}}_{\mathrm{I}}=1$), i.e. there is always an error regime where $P_{\mathrm{I}} < P^{\mathrm{Het}}_{\mathrm{I}}$. Ideally, this performance is achieved at the smallest possible value of $P_{\mathrm{I}}$, which will depend on the number of possible states (See Supplementary Note III).

 \textcolor{black}{We note that the proposed measurement uses similar techniques in the state elimination stage as the Bondurant receiver for MESD of multiple states \cite{bondurant93}, and the USD receiver based on feedback and state elimination \cite{banaszek99}. However, the proposed strategy in here makes use of the optimal inconclusive measurement of two states for allowing to transition between the measurement paradigms of MESD and USD to realize an optimized inconclusive measurement of multiple coherent states. While the performance of the strategy for more states will degrade due to the increased inconclusive probability in the state elimination stage, we expect that this strategy will serve as the basis for designing optimized inconclusive strategies in higher dimensions. A possible example for inconclusive measurement strategies could use hybrid measurement schemes combining Gaussian measurements, such as homodyne, with photon counting \cite{mueller12,chen18}. In these schemes, the Gaussian measurement can eliminate a sub-set of the states, and photon counting would be used for state elimination in a smaller sub-set of states followed by the binary optimal inconclusive measurement.}
\section{Discussion}
Optimal inconclusive measurements are generalized quantum measurements that encompass standard paradigms of state discrimination including MESD and USD. These non-projective measurements allow for diverse state discrimination tasks and provide a more powerful tool for \textcolor{black}{classical and quantum information processing \cite{wittmann10,sych10}. In optical communication, inconclusive measurement results can be treated as an erasure channel, and optimal inconclusive measurements can be leveraged to increase the amount of information transfer by utilizing communication codes well suited for erasure channels \cite{chen12}. These optimal inconclusive measurements can also enable hybrid repeater schemes where inconclusive measurements of coherent states are used to entangle remote quantum memories \cite{loock06,schmidt20}.} Recent advancements in quantum measurement theory showed that such complex quantum measurements for binary coherent states can be realized using \textcolor{black}{single photon detection} and local operations and classical communication in a two-mode measurement \cite{nakahira12}. This measurement strategy splits the energy of the input state into two temporal modes. It performs a MESD measurement of the input state in the first mode with a certain probability of error, and an optimal inconclusive measurement in the single-state domain in the second mode determining whether the measurement result is inconclusive. The optimality of this measurement makes it possible to achieve minimal error for a given inconclusive probability. Moreover, such generalized quantum measurements can be realized with Dolinar-\textit{like} optimal receivers for coherent states.

Here, we experimentally demonstrate the optimal inconclusive measurement proposed in \cite{nakahira12}. Our demonstration uses coherent displacement operations, single photon detection, and fast feedback to implement these general non-projective quantum measurements with high fidelity in a real system. We further use this measurement to demonstrate the optimal MESD for phase-encoded binary coherent states, which is the optimal modulation for optical communications under the average power constraint. \textcolor{black}{While our proof of principle demonstration of the optimal inconclusive measurement was realized at moderate measurement rates, future implementations based on integrated photonics with high-bandwidth optical modulation and processing \cite{esmaeilzadeh20} within small footprints, together with advancements in high-bandwidth integrated nanowire detectors will allow for demonstrations at GHz bandwidths.} These results show that Dolinar-\textit{like} receivers can be used to perform a wide variety of measurements within a two dimensional Hilbert space with current technologies. Furthermore, we show how the binary optimal inconclusive measurement can be leveraged to perform inconclusive measurements in higher dimensions with hybrid measurements using sequential unambiguous state elimination of multiple states. Our work contributes to our understanding of the fundamental and practical limits of measurements based on \textcolor{black}{single photon detection}, coherent displacement operations, and feedback, and can further our understanding of quantum measurement theory \cite{nakahira18}. Moreover, these measurement techniques can potentially allow for implementations of more general non-projective measurements in two dimensional spaces using linear optics and \textcolor{black}{single photon detection}.

\section{Methods}
\subsection{Optimal Displacement Waveform}
The optimal inconclusive measurement $\{\hat{\Pi}_{1}, \hat{\Pi}_{2}, \hat{\Pi}_{?}\}$ for binary coherent states can be realized with a generalized Dolinar receiver \cite{nakahira12} . In this modified strategy \cite{nakahira12}, the generalized receiver performs optimal measurements in two temporal modes during the measuring time $0 \leq t \leq 1$. In the first temporal mode, $0 \leq t \leq t_{1}$, the optimal inconclusive receiver performs an optimal MESD measurement to discriminate between $\{|\pm\alpha\rangle\}$ with minimal error using the optimal displacement waveform \cite{dolinar,cook07,geremia04b,nakahira12}:
\begin{equation}
u_{1}(0 \leq t \leq t_{1}) = \frac{(-1)^{N_{1}(t)}\alpha}{\sqrt{1 - 4p(1-p)K^{2t}}},
\label{dolinarwf}
\end{equation}
\noindent Here $K^{2} = |\langle -\alpha| \alpha \rangle|^{2} = e^{-4|\alpha|^{2}}$, $p$ is the prior probability of the most likely state, and $N_{1}(t)$ is the total number of detected photons in the first mode up to time $t\leq t_{1}$, where $N_{1}(0)=0$.
Note that during the first temporal mode the phase of the LO displacement field (sign of $u_{1}(t)$) switches between 0 and $\pi$ each time a photon is detected, similar to the Dolinar receiver \cite{dolinar76}. During the measurement in the first temporal mode, the provisional hypothesis for the input state at time $t$ is $|\alpha\rangle$ if $N_{1}(t)$ is even and $|-\alpha\rangle$ if $N_{1}(t)$ is odd. The provisional probabilities for the two input states after the first temporal mode are $\{P_{\mathrm{C}}^{(1)}, 1-P_{\mathrm{C}}^{(1)}\}$ with:
\begin{equation}
P_{\mathrm{C}}^{(1)} = \frac{1}{2} \Big(1 - \sqrt{1 - p(1-p) e^{-4t_{1}|\alpha|^{2}}} \Big),
\end{equation}
\noindent which corresponds to the Helstrom bound for the coherent states $\{|\pm\sqrt{t_{1}}\alpha\rangle\}$. The optimal waveform $u_{1}(t)$ in Eq. (\ref{dolinarwf}) and the evolution of t	he probability of correct detection $P_{\mathrm{C}}(t)$ at $t$ can be obtained using Bayesian updating \cite{assalini11,acin05,nakahira12} or optimal control \cite{geremia04b}.

In the second temporal mode ($t_{1}<t\leq1$), the receiver performs an optimal inconclusive measurement in the so-called single-state domain, where $P_{\mathrm{I}}$, $P_{\mathrm{C}}^{(1)}$, and $(1-t_{1})|\alpha|^{2}$ are such that $\hat{\Pi}_{2}=0$ \cite{nakahira12,sugimoto09}. Without loss of generality, the most probable state is $|\alpha\rangle$, and the non-zero POVM elements are $\hat{\Pi}_{1}$ and $\hat{\Pi}_{?}$. This optimal inconclusive projective measurement in the single-state domain can be realized by a Dolinar-\textit{like} receiver with optimal displacement waveform \cite{nakahira12}:
\begin{equation}
u_{2}(t_{1}<t\leq1) = \frac{(-1)^{N_{2}(t)+N_{0}}\alpha}{\sqrt{1 - 4v(1-v)K^{2t}}},
\label{dolinarwf2}
\end{equation}
\noindent where $p$ in Eq. (\ref{dolinarwf}) is replaced by the quantity $v$, which depends on $P_{\mathrm{I}}$, $p$, and $|\alpha|^{2}$ \cite{nakahira12}. $N_{2}(t)$ is the number of photons detected in the second mode with $N_{2}(t_{1})=0$, and $N_{0}$ determines the phase of the LO at $t_{1}$: $N_{0}=0$ if $v > 0.5$ and $N_{0}=1$ otherwise.

The total displacement waveform for the optimal inconclusive receiver is thus a combination of $u_{1}(t)$ in Eq. (\ref{dolinarwf}) and $u_{2}(t)$ in Eq. (\ref{dolinarwf2}) resulting on the total optimal displacement in Eq. (\ref{totalwf}) in main text. This strategy therefore implements a standard Dolinar receiver during the first mode $0 \leq t \leq t_{1}$, and then a Dolinar-\textit{like} receiver during $t_{1} < t \leq 1$ assuming the input states at $t=t_{1}$ have prior probabilities $\{v, 1-v\}$ \cite{nakahira12}. %

\subsection{Experimental Setup Details}
In our experimental demonstration, optical pulses are generated from a Helium-Neon laser and a pulsed acousto-optic modulator (AOM), and then split into the signal arm (upper) and LO arm (lower), as shown in Figure \ref{optincth}(d). The input states are prepared with an attenuator (Att.) and a phase modulator (PM). The LO field is prepared by a PM with a multiplexer (MUX), and an amplitude modulator (AM) with a digital-to-analog converter (DAC). The input state and the LO field interfere on a 99/1 beam splitter (BS) to implement the optimal displacement waveforms $\hat{D}(u(t))$ conditioned on photon detection  events using a single photon detector (SPD). A field programmable gate array (FPGA) stores the magnitude of the optimal waveform $|u(t)|$ in Eq. (\ref{totalwf}) in memory, prepares the amplitude and phase of the LO conditioned on $N_{1}(t)$, $N_{2}(t)$, and $N_{0}$, and implements the strategy for the optimal inconclusive measurement. We discretize time $t$ into 1024 time bins of 160 ns each where a photon can be detected to approximate a continuous measurement. \textcolor{black}{Our implementation achieves a feedback bandwidth of about 6 MHz, which is limited by the APD output latency, electronic bandwidth of controllers, switches and FPGA, accounting for about 50 ns and optical delays in the interferometric setup (100 ns).} The FPGA processes and stores the photon detections during these time bins and sends the detection histories to a computer. We reconstruct the measurement probabilities in post-processing. The optimal inconclusive measurement requires very large values of the ratio between mean photon numbers of the displacement field and the input state, $R=|u(t)|^{2}/|\alpha|^{2}$. However, experimentally there is a maximum ratio $R$ that can be reliably implemented. In our demonstration, we set the maximum of this ratio to $R=50$, which is limited by the extinction ratio ($\approx 20$ dB) of the AM in the LO arm of the setup. The impact of finite values for $R$ and other experimental imperfections are discussed in the Supplementary Notes I \& II.

\subsection{FPGA Implementation}
\textcolor{black}{We use an Opal Kelly ZEM4310 to control the experiment, which is based on an Altera Cyclone IV FPGA and has a base clock rate of 50 MHz. We discretize the discrimination measurements into 1024 time bins of 160 ns each such that a single shot of the experiment corresponds to a pulse that is 163.8 $\mu$s long. The magnitude of the LO waveform for each of the 1024 time bins is pre-calculated for each $|\alpha|^{2}$ and inconclusive probability $P_{\mathrm{I}}$ and stored in a look-up table within the FPGA as an 8-bit value. The phase of the LO flips between 0 and $\pi$ each time a photon is detected. This method for preparing the optimal LO waveform given by Eq. (1) allows us to efficiently implement the desired optimal inconclusive measurement in our demonstration.}

\textcolor{black}{The optimal inconclusive strategy requires precise and fast control of the LO phase. We control the phase of the LO by changing the voltage applied to the phase modulator between two values which correspond to a phase of 0 and a phase of $\pi$. Each time the phase of the LO changes, we ignore the output of the APD for 160 ns to avoid accidental photon detections. This “blanking time” is obtained by noting that the combined electrical and optical delay time between changing the modulation voltage and observing the corresponding photons at the APD is approximately 150 ns. This also has the benefit of reducing the effective after-pulsing probability to close to zero. Typically, the probability of detecting an after-pulse is at its maximum immediately after the dead-time of the APD ($\approx40$ ns for our implementation), but this probability quickly decays with time. We note that without any blanking, the cumulative after-pulsing probability of our APD is $P_{\mathrm{AP}}\approx 0.015$ and $P_{\mathrm{AP}} < 0.001$ with 100 ns of blanking}

\subsection{Interferometer Stabilization}
\textcolor{black}{In order to maintain a well-defined relative phase between the signal and LO fields, we actively stabilize the interferometer. We run the experiments with a 4 kHz repetition rate to give an experimental duty cycle of $\approx 66\%$ (experiment time of $\approx$ 165 $\mu$s, locking time of $\approx$ 91 $\mu$s). During the part of the experimental duty cycle when the experiment is not taking place, the relative phase between the two arms of the interferometric setup actively stabilized with a feedback loop using a PID controller and a piezo on the back of a mirror in the signal arm, see Fig. 2. We obtain the error signal for stabilization of the interferometer using a narrow-band laser at 780 nm, which is actively stabilized in frequency to an atomic line in rubidium using saturated absorption spectroscopy. Light from this laser propagates in an opposite direction through the interferometer compared to the light at 633 nm, and is detected with a differential detector to measure the phase fluctuations. Slightly before the discrimination measurement begins, the feedback loop is paused and voltage to the piezo is fixed at its current value. The stabilization feedback loop resumes after the discrimination measurement is completed.}

\subsection{Optical Modulators}
\textcolor{black}{The setup uses fiber coupled Lithium-Niobate, amplitude and phase electro-optic modulators (AM and PM), with a 3 dB bandwidth of $\approx1$ GHz. The phase modulators (PMs) have a $\pi$-voltage of $V_{\pi}=1.5$ V and the amplitude modulator (AM) has a $\pi$-voltage of $V_{\pi} = 750$ mV with an extinction ratio of $\approx 20$ dB. The amplitude and phase of the LO and the phase of the signal fields are adjusted using three 8-bit digital-to-analog converters (DAC), voltage-controlled gain circuits, and summing amplifiers.}
\\
\\
\noindent
\textbf{ACKNOWLEDGEMENTS}
\\
This work was supported by the National Science Foundation (NSF) (PHY-1653670).
\\
\\
\noindent
\textbf{AUTHOR CONTRIBUTIONS}
\\
F.E.B. supervised the work. M.T.D. designed the experimental implementation and performed the measurements. All authors contributed to the analysis of the theoretical and experimental results, conceived the idea of generalizing inconclusive measurement to higher dimensions, and contributed to writing the manuscript.
\\
\\
\noindent
\textbf{COMPETING INTERESTS}
\\
The authors declare that there are no competing interests.
\\
\\
\textbf{DATA AVAILABILITY}
\\
The data that support the findings of this study are available from the authors upon
request.
\\
\\
\noindent

%

\makeatletter
\renewcommand{\fnum@figure}{~\textbf{Supplementary Figure \thefigure}}
\makeatother

\newpage

\onecolumngrid

\begin{center}
	\textbf{\large Supplementary Material: Demonstration of Optimal Non-Projective Measurements of Binary Coherent States with Photon Counting}
\end{center}

\twocolumngrid
\section*{Supplementary Note I: Impact of Experimental Imperfections}
\begin{figure}[t]
	\includegraphics[width = 8.5cm]{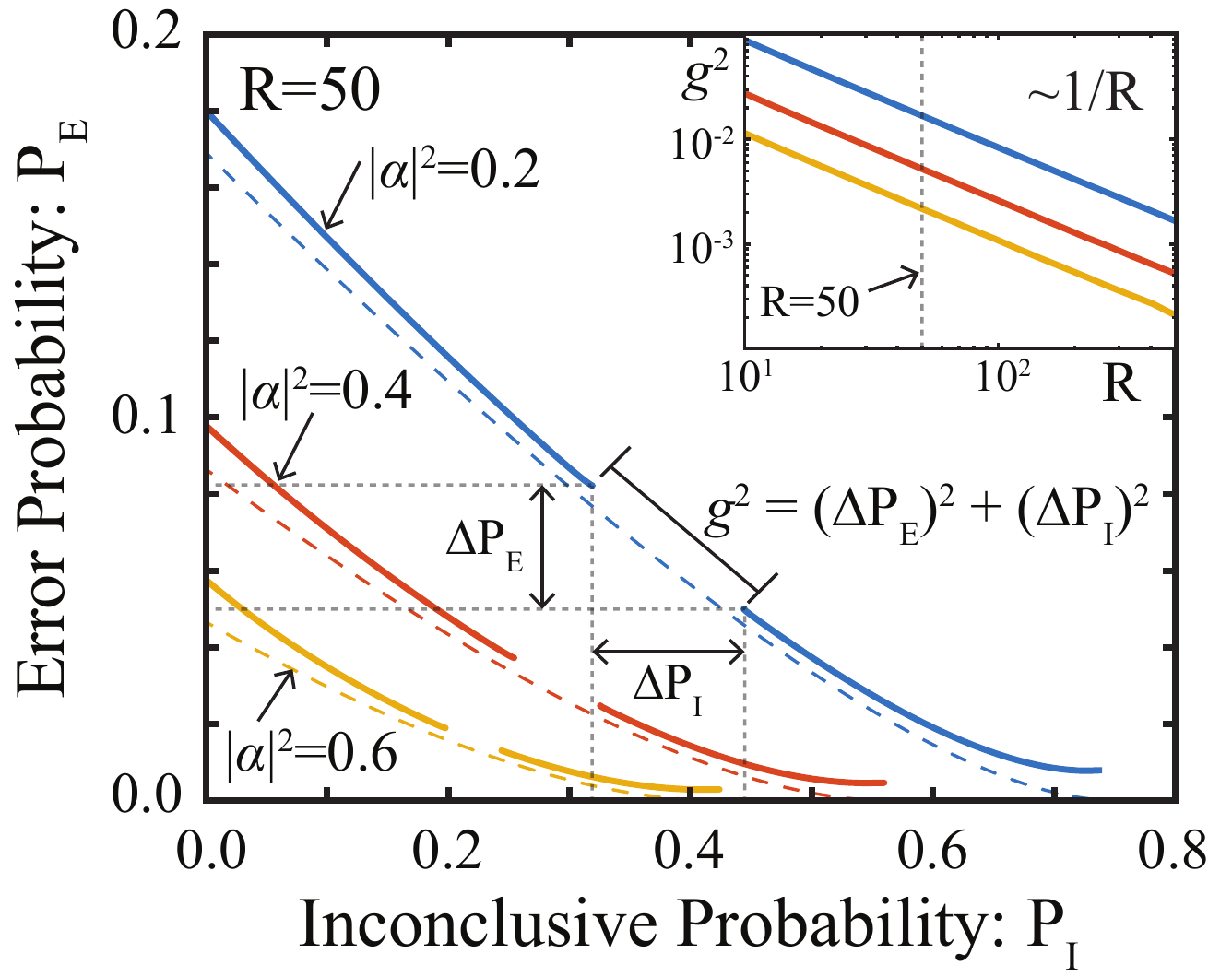}
	\caption{Impact of a finite value of $R=\mathrm{max}(|u(t)|^{2})/|\alpha|^{2}$ on the optimal inconclusive measurement. The blue, orange, and yellow solid lines show the simulated results for $|\alpha|^{2}$=0.2, 0.4, and 0.6, respectively, for a strategy with $R=50$, and the dashed lines show the results for $R=\infty$. A value of $R<\infty$ produces a gap $g^2 = (\Delta P_{\mathrm{E}})^{2} + (\Delta P_{\mathrm{I}})^{2}$ in the $P_{\mathrm{I}}-P_{\mathrm{E}}$ curve. The dashed gray lines show the ranges $\Delta P_{\mathrm{I}}$ and $\Delta P_{\mathrm{E}}$ for $|\alpha|^{2}$=0.2 that cannot be accessed by the measurement. The inset shows the scaling of the gap $g$ as a function of $R$, which follows $~1/R$ scaling for each value of $|\alpha|^{2}$.}
	\label{gap}
\end{figure}

We investigate the effect of non-ideal experimental parameters on the performance of the optimal inconclusive measurement and the Dolinar receiver. A critical experimental parameter is the ratio $R=\mathrm{max}(|u(t)|^{2})/|\alpha|^{2}$ between the maximum local oscillator (LO) energy $\mathrm{max}(|u(t)|^{2})$ and the energy of the input state $|\alpha|^{2}$ that can be reliably implemented in the experiment. In principle, the optimal inconclusive measurement requires very large values of $R$ (e.g. $R=\infty$ at $t=0$ for equiprobable states). However, having a finite value of $R$ that can be applied limits our ability to implement the desired optimal displacement $u(t)$ perfectly, which impacts the performance of the strategy. Supplementary Figure \ref{gap} shows the effect of a finite value of $R=50$ on the optimal inconclusive measurement for $|\alpha|^{2}$=0.2, 0.4, and 0.6. The dashed lines show the performance for when $R=\infty$, for reference. We obtain each data point by solving the evolution equations for $P_{\mathrm{I}}$ and $P_{\mathrm{E}}$ using the bounded optimal waveform $u'(t) = \mathrm{min}[u(t), \sqrt{R}|\alpha|\mathrm{sign}(u(t))]$ with a finite $R$, and then plotting the obtained values of $P_{\mathrm{E}}$ and $P_{\mathrm{I}}$. We observe that the main effect of finite $R$ is a gap $g$ in the achievable $P_{\mathrm{I}}-P_{\mathrm{E}}$ curve. We define this gap as $g^2 = (\Delta P_{\mathrm{E}})^{2} + (\Delta P_{\mathrm{I}})^{2}$, where the dashed gray lines show $\Delta P_{\mathrm{E}}$ and $\Delta P_{\mathrm{I}}$ for $|\alpha|^{2}=0.2$. The inset shows $g^2$ as a function of $R$ for each value of $|\alpha|^{2}$ on a log-log scale, showing a $1/R$ scaling.

\begin{figure*}[t]
	\includegraphics[width = \textwidth]{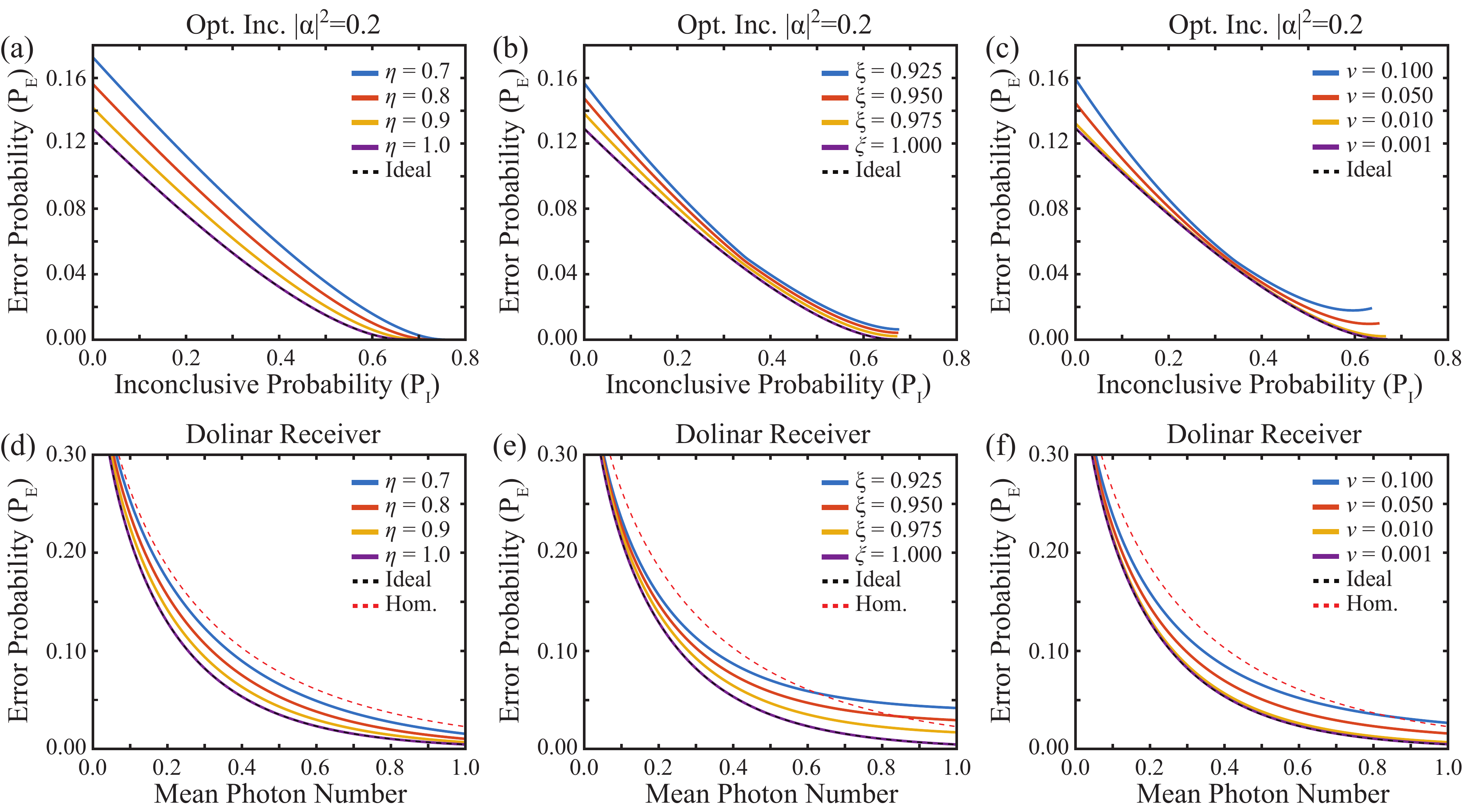}
	\caption{Impact of imperfections on the optimal inconclusive measurement (top row) and the optimal MESD measurement, the Dolinar receiver (bottom row). Colored lines show the effect of different levels of detection efficiency $\eta$ (a,d), interference visibility $\xi$ (b,e), and dark counts $\nu$ (c,f) on the attainable performances of the corresponding measurements for $|\alpha|^{2}=0.2$.
	}
	\label{exp_study}
\end{figure*}

Intrinsic imperfections such as non-ideal interference visibility and non-zero dark counts can produce unexpected photon detections, which cause errors in the discrimination process. Supplementary Figure \ref{exp_study}(a-c) shows the performance of the optimal inconclusive measurement for $|\alpha|^{2}=0.2$ for different levels of experimental imperfections (solid colored lines) compared to the ideal case (dashed line). Non-ideal values of (a) the detection efficiency $\eta$, (b) the interference visibility $\xi$, and (c) the dark count rate $\nu$ all increase the attainable error probability compared to the ideal case. A reduction in $\eta$ increases the error probability by approximately the same amount across the entire range of values of $P_{\mathrm{I}}$. On the other hand, higher dark counts have a larger impact closer to the MESD ($P_{\mathrm{I}}\approx 0$) and USD ($P_{\mathrm{E}} \approx 0$) regimes. Supplementary Figure \ref{exp_study}(d-f) shows the performance of the optimal MESD measurement, the Dolinar receiver, for different levels of experimental imperfections (solid colored lines) compared to the ideal case (dashed black line) and homodyne limit (dashed red line). We observe that for the Dolinar receiver, non-ideal interference visibility (e) has the largest detrimental effect, significantly increasing the error probability at larger values of $|\alpha|^{2}$.
\section*{Supplementary Note II: Phase Correction Calibration}

The LO amplitude modulator (AM) also incidentally modulates the LO phase depending on the depth of the amplitude modulation, which is related to the parameter $R$. The relationship between the applied amplitude modulation and the resulting extra phase modulation is typically linear for low modulation depths. However, the implementation of the optimal inconclusive measurement requires modulation as large as experimentally possible (in principle infinitely large at $t=0$ for equal prior probabilities) to implement the necessary displacement waveform for a particular strategy. Thus, to correct the phase modulation caused by the AM we apply a phase correction to the phase modulator (PM) in the LO arm, such that any extra phase modulation from the AM will be exactly canceled out once the light passes through both modulators. This correction allows us to experimentally implement the displacement waveform which implements the desired optimal inconclusive measurement. A critical aspect of the strategy for the optimal inconclusive measurement is that the optimal waveform for the LO magnitude has a deterministic shape, i.e. takes on values which implement a predetermined displacement magnitude. Since the LO magnitude is predetermined, the necessary phase correction is predetermined as well. Thus, in our implementation of this strategy in the FPGA, the correction for the unwanted phase shift introduced by the amplitude modulator is included as an extra factor for the phase of the LO in the FPGA. This modification allows us to realize feedforward operations to the PM of the LO in real time for implementing the phase correction of this undesirable effect.

We calibrate this phase correction by measuring the voltage required to correct the phase shift from the AM as a function of the bit-value (proportional to voltage via the DAC) applied to the AM. To measure the necessary amount of phase correction, we use an interferometric measurement where the Signal and LO fields interfere, while the interferometer is stabilized. We apply two copies of a waveform for phase calibration to the PM for the Signal field back-to-back within a single light pulse. The phase calibration waveform implements the phase pattern [0, $\pi/2$, 0, $\pi$, 0, $3\pi/2$, 0], where each value has the same time duration \cite{becerra13}. During the first copy, the phase and amplitude modulation of the LO is turned off, and during the second copy we enable the AM of the LO with a particular bit-value.

We calibrate the phase shift induced by the AM of the LO by observing the intensity of the interference pattern after the Signal and LO interfere. Comparing the measured intensity for relative phases $\pi/2$ and $3\pi/2$ allows us correct the phase shift induced by the AM by adjusting the PM of the LO to achieve the expected interference for these two phases. These two phases should correspond to the same intensity, but the phase error from the AM causes them to be uneven and we simply adjust the phase correction until they are balanced. We repeat this process to obtain the phase correction voltage for different values of amplitude modulation to obtain a phase correction calibration curve. This calibration curve corresponds to the extra voltage we must apply to the LO PM to cancel out the phase error from amplitude modulation in the measurement. During the experimental runs, this phase correction is simply added to the current LO phase, which depends on the random photon detections in the measurement, and the strength of the correction depends on the value sent to the AM of the LO, which is coded in the FPGA as a look-up table.

\begin{figure}[t]
	\includegraphics[width = 8.5cm]{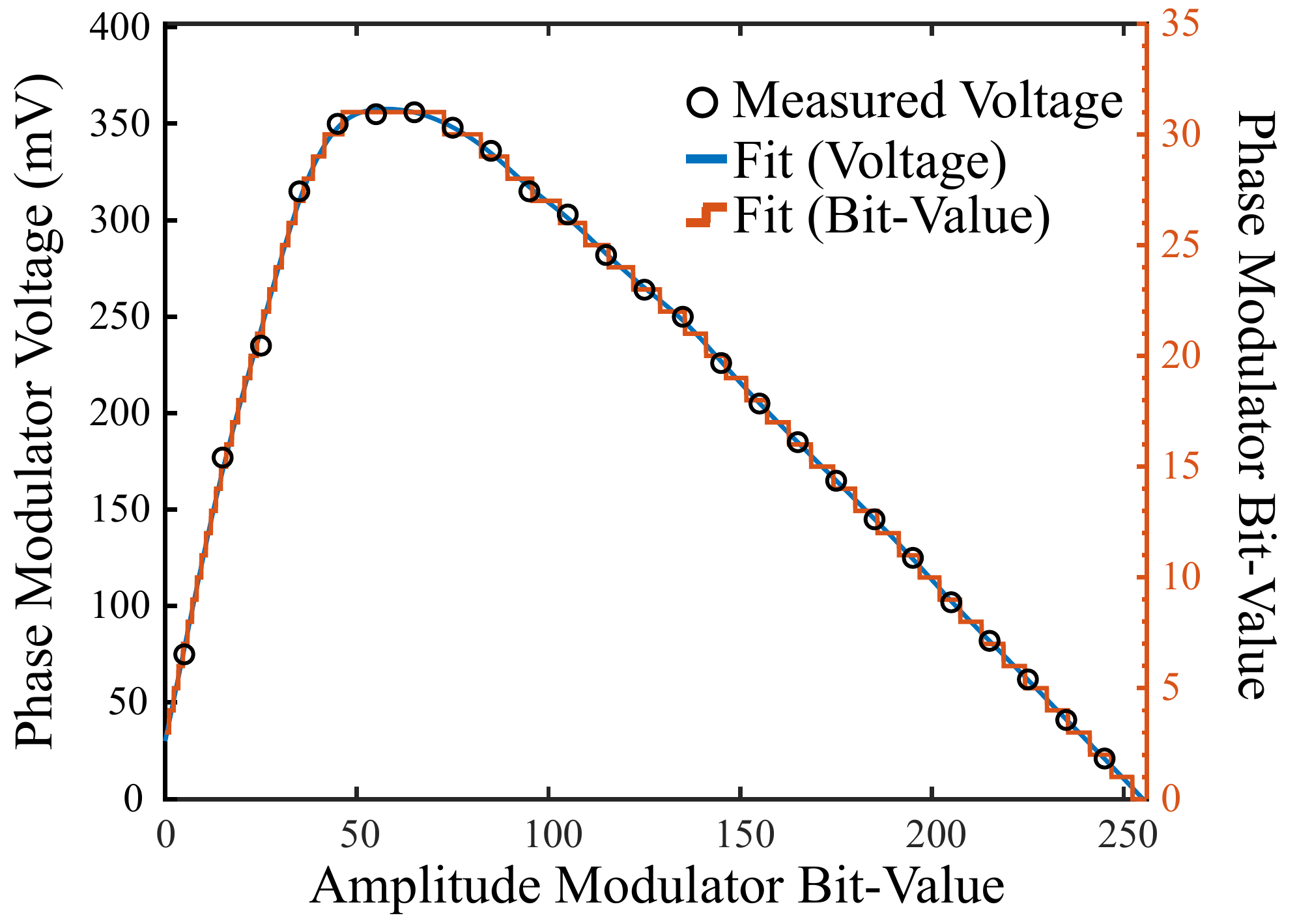}
	\caption{Example correction curve for the phase error imparted by the LO AM. The points show the phase modulator voltage required to correct the incidental phase shift from amplitude modulation. At low modulation depth (large bit-value due to our configuration), the correction is relatively linear as expected. However at large modulation (small bit-value) the needed phase correction becomes highly non-linear. The blue line shows the data fit using a smoothing spline and the orange line shows the corresponding bit-value added to the phase of the LO to correct the phase error.}
	\label{amcal}
\end{figure}
Supplementary Figure \ref{amcal} shows an example of the measured phase correction curve. The dots show the measured voltage applied to the PM required to correct the phase error from amplitude modulation. Different points on the \textit{x}-axis (Amplitude Modulation Bit-Value) correspond to different levels of amplitude modulation depth which implement different displacement magnitudes. The blue line shows a fit to the measured points using a smoothing spline. The orange curve shows the bit-value that needs to be applied to the phase modulator in order to correct the phase error, which appears to be piece-wise constant due to the finite precision in digitization (8-bits) within the FPGA. For example, an amplitude modulation bit-value of 150 will require the FPGA to add a bit-value of 19 to the LO phase to cancel out any phase errors, and an amplitude modulation value of 50 requires a phase correction bit value of 31 be added to the LO phase. This correction ensures that there is no incidental phase shifts caused by amplitude modulation, allowing us to precisely implement the correct displacement waveform needed for a particular optimal inconclusive measurement.

\section*{Supplementary Note III: Extending to High Dimensional Inconclusive Measurements}
\textcolor{black}{We show how to construct inconclusive measurements in high dimensions using the combination of hypothesis testing followed by the binary optimal inconclusive measurement. We extend the strategy proposed in Section III to M-PSK coherent state alphabets by again dividing the single measurement into two stages. In the first stage, the receiver implements hypothesis tests until there are only two possible remaining states. There are on the order of $2^M$ possible detection histories for the hypothesis testing stage, and most of these outcomes will result in an inconclusive result, which yields $P^{(1)}_{\mathrm{I}}$. The binary optimal inconclusive measurement in the second stage can then achieve arbitrarily small error probabilities $P_{\mathrm{E}}$ at the expense of increasing the total inconclusive probability $P_{\mathrm{I}}$ by a relatively small amount given by $P^{(2)}_{\mathrm{I}}$. In contrast, heterodyne detection requires that $P^{\mathrm{Het}}_{\mathrm{I}}$ asymptotically approaches $P^{\mathrm{Het}}_{\mathrm{I}}=1$ in order to reach asymptotically small error probabilities $P^{\mathrm{Het}}_{\mathrm{E}}$ (though $P^{\mathrm{Het}}_{\mathrm{E}}=0$ only at $P^{\mathrm{Het}}_{\mathrm{I}}=1$). Thus, examining the minimum possible inconclusive probability $P^{(1)}_{\mathrm{I}}$ allows us to examine the scaling in the performance of the proposed hybrid approach as the number of states $M$ increases, since lower error rates can always be achieved by a small increase in the total $P_{\mathrm{I}} = P^{(1)}_{\mathrm{I}} + P^{(2)}_{\mathrm{I}}$ where $P_{\mathrm{I}} < 1$ for $P_{\mathrm{E}} = 0$.}

\textcolor{black}{The minimum inconclusive probability $P^{(1)}_{\mathrm{I}}$ is given by the total probability that the hypothesis testing stage results in a detection history where there are three or more possible states remaining. A given state $s_{j} \in \{s_{1}...s_{M}\}$ is eliminated as a possible state if the receiver detects a photon when testing for that state. The entire hypothesis testing stage consists of testing for each possible input state until either only two states remain, or all states have been tested. We denote the detection history of the hypothesis testing stage as $\textbf{d}_{i}$ and the set of all detection histories which result in an inconclusive result as $\textbf{D}_{\mathrm{inc}}$. We then obtain the minimum inconclusive probability $P^{(1)}_{\mathrm{I}}$ as:}

\begin{align}
	P^{(1)}_{\mathrm{I}} &= \sum_{\textbf{d}_{i}\in\textbf{D}_{\mathrm{inc}}} P(\textbf{d}_{i}) 
	\\
	&= \sum_{\textbf{d}_{i}\in\textbf{D}_{\mathrm{inc}}} \sum_{j=1}^{M} P(\textbf{d}_{i} | s^{*} = s_{j})P(s^{*} = s_{j})
	\\
	&= \sum_{\textbf{d}_{i}\in\textbf{D}_{\mathrm{inc}}} \sum_{j=1}^{M} \prod_{k=1}^{M} \mathcal{L}(d_{i, k} | \hat{s} = s_{k}, s^{*} = s_{j}) \times \frac{1}{M}
\end{align}

\noindent \textcolor{black}{where $P(\textbf{d}_{i})$ is the probability of obtaining detection history $\textbf{d}_{i}$ from the hypothesis testing stage, $P(\textbf{d}_{i} | s^{*} = s_{j})$ is the probability of obtaining the detection history $\textbf{d}_{i}$ given the actual input state $s^{*}$ is $s_{j}$, and $P(s_{j})=1/M$ is the prior probability. Also, $\mathcal{L}(d_{i, k} | \hat{s} = s_{k}, s^{*} = s_{j})$ is the photon counting likelihood of detecting $d_{i, k}$ from a single hypothesis test given that the receiver tests the state $\hat{s} = s_{k}$ and the actual input state is $s^{*} = s_{j}$. We note that $\mathcal{L}(1 | s_{j}, s_{j}) = 0$ and this enforces that some detection histories never actually occur. Given that $P^{(1)}_{\mathrm{I}} \rightarrow 1$ as $M \rightarrow \infty$, we obtain the maximum \textit{conclusive} probability as $P^{\mathrm{max}}_{\mathrm{conc}} = 1 - P^{(1)}_{\mathrm{I}}$. }

\textcolor{black}{As an example, the minimum inconclusive probability $P^{(1)}_{\mathrm{I}}$ for $M=3$ as in Sec. III of the main text is given by the probability of detecting zero photons in each hypothesis test, i.e. obtaining $\textbf{d}_{0}=\{0,0,0\}$. The histories $\textbf{d}_{1}=\{0, 0, 1\}$, $\textbf{d}_{2}=\{0, 1, 0\}$, and $\textbf{d}_{4}=\{1, 0, 0\}$ result in performing the binary optimal inconclusive measurement on the remaining two states, and the other detection histories never occur. Thus, $\textbf{D}_{\mathrm{inc}}=\{\textbf{d}_{0}\}$ and we obtain:}
\begin{align}	
	P^{(1)}_{\mathrm{I}} &= P(\textbf{d}_{0}) \\
	\nonumber
	&= \sum_{j=1}^{3} \prod_{k=1}^{3} \mathcal{L}(0 | s_{j}, s_{k})/3 \\
	\nonumber
	&= (1) \Bigg(e^{-2 \frac{f_{3}|\alpha|^{2}}{3}\big[1 - \mathrm{cos}(\frac{2\pi}{3})\big]} \Bigg)  \Bigg(e^{-2 \frac{f_{3}|\alpha|^{2}}{3}\big[1 - \mathrm{cos}(\frac{4\pi}{3})\big]} \Bigg).
\end{align}

\noindent \textcolor{black}{Which yields $P_{\mathrm{I}}^{(1)} \approx \{0.766, 0.587, 0.449\}$ for $|\alpha|^{2} = \{0.2, 0.4, 0.6\}$ respectively with $f_{3}=2/3$ as in Sec. III of the main text.}

\begin{figure}[t]
	\includegraphics[width = 8.5cm]{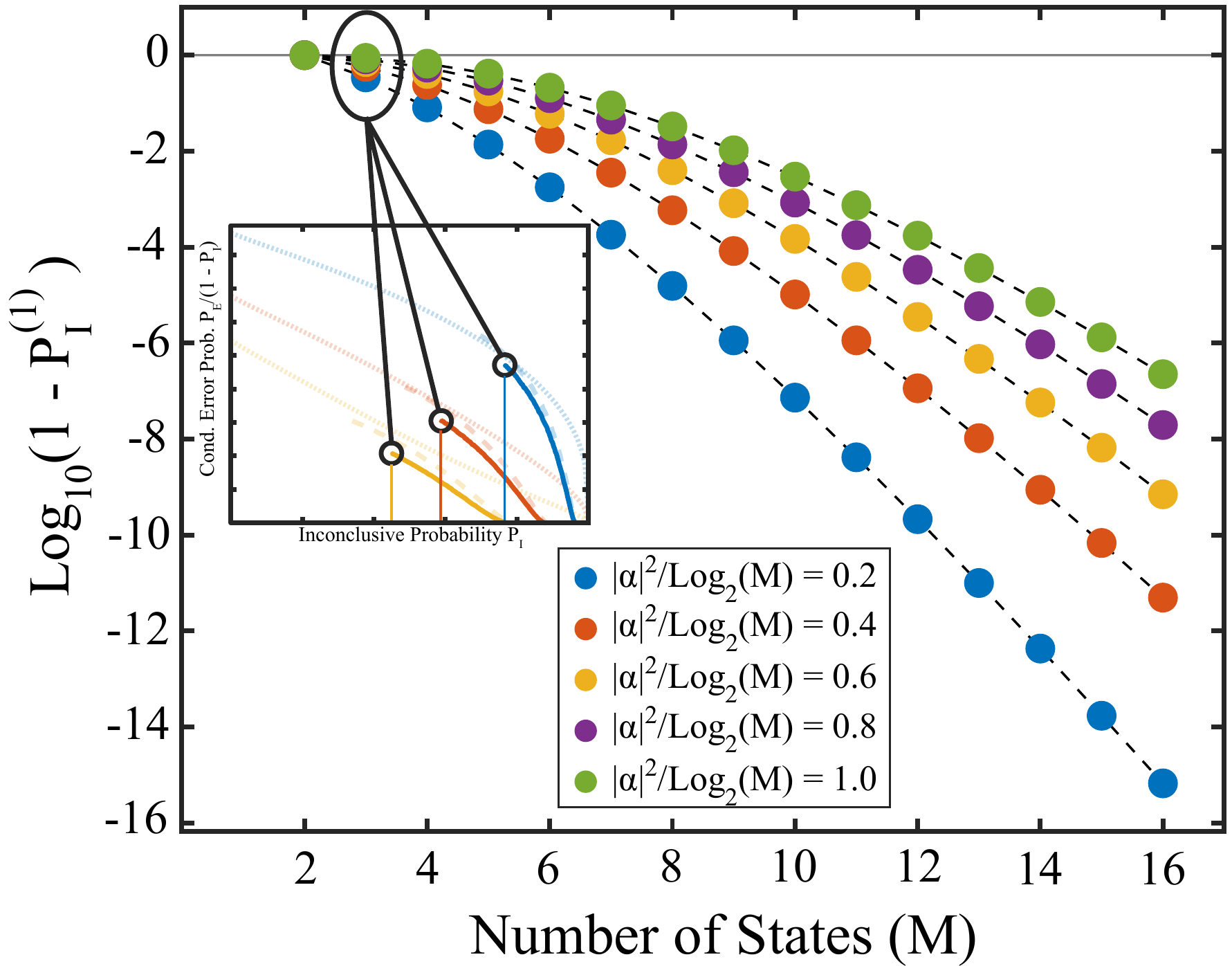}
	\caption{\textcolor{black}{Scaling of the maximum conclusive probability for inconclusive measurements based on hypothesis testing for high dimensional M-PSK alphabets. Colored points show the maximum conclusive probability $P^{\mathrm{max}}_{\mathrm{conc}} = 1 - P^{(1)}_{\mathrm{I}}$, where a larger $P^{\mathrm{max}}_{\mathrm{conc}}$ is better, for different numbers of states in the PSK alphabet. The value of $P^{\mathrm{max}}_{\mathrm{conc}}$ increases as the mean photon number per bit $|\alpha|^{2}/\mathrm{log}_{2}(M)$ increases and $\mathrm{log}_{10}(1 - P^{(1)}_{\mathrm{I}})$ scales approximately proportional to $M^{2}$.}}
	\label{mpsk_oim}
\end{figure}

\textcolor{black}{Supplementary Figure \ref{mpsk_oim} shows the scaling of the maximum conclusive probability $\mathrm{log}_{10}(P^{\mathrm{max}}_{\mathrm{conc}}) = \mathrm{log}_{10}(1 - P^{(1)}_{\mathrm{I}})$ as the number of states $M$ in the alphabet increases. The blue, orange, yellow, purple, and green points show the results for a mean photon number per bit of $|\alpha|^{2}/\mathrm{log}_{2}(M) = $ 0.2, 0.4, 0.6, 0.8, and 1.0. The dashed lines are to guide the eye. We note that here the fraction $f$ of total energy $|\alpha|^{2}$ used to perform the hypothesis testing changes for each alphabet as $f_{M} = (M-1)/M$ such that here $M=3$ (inset) corresponds to the results shown in Supplementary Fig. (3) in the main text. We find that the quantity $\mathrm{log}_{10}(1 - P^{(1)}_{\mathrm{I}})$ roughly scales proportionally to $M^{2}$ for all total mean photon numbers, suggesting that $P^{\mathrm{max}}_{\mathrm{conc}} \propto 10^{c_{\alpha}\times M^2}$ where the coefficient $c_{\alpha}<0$ is specific to a particular value of $|\alpha|^{2}$. We note that this scaling may change with a different choice of $f_{M}$. Furthermore, for $M \geq 4$, a Bayesian approach to hypothesis testing may provide better performance as opposed to randomly choosing between the remaining untested states, such as in Refs. \cite{becerra13, ferdinand17, burenkov20}.}
\section*{Supplementary Note IV: Optimal Inconclusive Measurement Simulation Details}

We simulate the optimal inconclusive measurement using two methods, one based on evolving equations derived in Ref. \cite{nakahira12} and the other based on Monte-Carlo sampling. Both methods allow for inclusion of experimental imperfections such as detection efficiency $\eta$, interference visibility $\xi$, dark count rate $\nu$, finite maximum displacement power $R$, and finite digitization precision from the digital-to-analog conversion within the experiment. A particular optimal inconclusive measurement is fixed by first setting the input energy $|\alpha|^{2}$, desired inconclusive probability $P_{\mathrm{I}}$, and prior probabilities of the two states $\{p, 1-p\}$. These parameters allow for calculation of the switching time $t_{1}=t_{1}(|\alpha|^{2}, P_{\mathrm{I}}, p)$ as well as the value of $v = v(|\alpha|^{2}, P_{\mathrm{I}}, p)$, which determines the phase of the displacement field at the beginning of the second temporal mode \cite{nakahira12}.

\begin{figure}[t]
	\begin{algorithm}[H]
		\caption{Optimal Inconclusive Simulation: Equations}
		\begin{algorithmic}[1]
			\scriptsize
			\State Fix: $|\alpha|^{2}$, $P_{\mathrm{I}}$, and $p$ \Comment{Energy, inconclusive probability, and prior probability of more likely input state}
			\State Fix: $\eta$, $\xi$, $\nu$, $R$ \Comment{Experimental Imperfections}
			\State
			\State $t_{1} \leftarrow t_{1}(|\alpha|^{2}, P_{\mathrm{I}}, p)$ \Comment{Eq. 13 in Ref. \cite{nakahira12}}
			\State $v \leftarrow v(|\alpha|^{2}, P_{\mathrm{I}}, p)$ \Comment{Eq. 18 in Ref. \cite{nakahira12}}
			\State
			\State Get $|u(t)|$ from Eq. (1)
			\State $|u_{\mathrm{exp}}(t)| = \mathrm{min}(\sqrt{R}, |u(t)|)$
			\State Transform $|u_{\mathrm{exp}}(t)|$ to nearest bit-value
			\State Transform $|u_{\mathrm{exp}}(t)|$ back to amplitude
			\State
			\State \textbf{Initial:} $P_{\mathrm{C}}(0) = p$, $P_{\mathrm{I}}(0) = 0$, $t = 0$
			\State $P_{\mathrm{E}}(0) = 1 - P_{\mathrm{C}}(0) - P_{\mathrm{I}}(0)$
			\While{$0 \leq t < t_{1}$} \Comment{Evolve for first temporal mode}
			\State $|\beta| \leftarrow |u_{\mathrm{exp}}(t)|$
			\State $\langle \hat{n} \rangle_{\pm} \leftarrow  \eta (|\alpha|^{2} + |\beta|^{2} \pm 2\xi |\beta||\alpha| ) + \nu $
			\State $P_{\mathrm{C}}(t + \delta t) \leftarrow P_{\mathrm{C}}(t) \big( 1 - \delta t\langle \hat{n} \rangle_{-} \big) + \big(1 - P_{\mathrm{C}}(t) \big) \delta t\langle \hat{n} \rangle_{+}$
			\State $P_{\mathrm{E}}(t + \delta t) \leftarrow 1 - P_{\mathrm{C}}(t + \delta t)$
			\State $P_{\mathrm{I}}(t + \delta t) \leftarrow 1 - P_{\mathrm{C}}(t + \delta t) - P_{\mathrm{E}}(t + \delta t)$
			\State $t \leftarrow t + \delta t$
			\EndWhile \Comment{End of first temporal mode}
			\State
			\If{$v \leq 0.5$}
			\State $P_{\mathrm{C}}(t_{1}) \leftarrow 0$
			\State $P_{\mathrm{I}}(t_{1}) \leftarrow 1$
			\State $P_{\mathrm{E}}(t_{1}) \leftarrow 1 - P_{\mathrm{C}}(t_{1}) - P_{\mathrm{I}}(t_{1})$
			\ElsIf{$v > 0.5$}
			\State $P_{\mathrm{C}}(t_{1}) \leftarrow P_{\mathrm{C}}(t_{1})$
			\State $P_{\mathrm{I}}(t_{1}) \leftarrow P_{\mathrm{I}}(t_{1})$
			\State $P_{\mathrm{E}}(t_{1}) \leftarrow 1 - P_{\mathrm{C}}(t_{1}) - P_{\mathrm{I}}(t_{1})$
			\EndIf
			\State
			\State $p' \leftarrow \frac{1}{2} \big( 1 - \sqrt{1 - 4p(1-p)e^{-4t_{1}|\alpha|^{2}}} \big)$ \Comment{Helstrom Bound for states  $| \sqrt{t_{1}}|\alpha| \rangle$}
			\While{$t_{1} \leq t < 1$} \Comment{Evolve for second temporal mode}
			\State $|\beta| \leftarrow |u_{\mathrm{exp}}(t)|$
			\State $\langle \hat{n} \rangle_{\pm} \leftarrow  \eta (|\alpha|^{2} + |\beta|^{2} \pm 2\xi |\beta||\alpha| ) + \nu $
			\State $P_{\mathrm{C}}(t + \delta t) \leftarrow P_{\mathrm{C}}(t) \big( 1 - \delta t\langle \hat{n} \rangle_{-} \big) + \big(p' - P_{\mathrm{C}}(t) \big) \delta t\langle \hat{n} \rangle_{+}$
			\State $P_{\mathrm{E}}(t + \delta t) \leftarrow P_{\mathrm{E}}(t) \big( 1 - \delta t\langle \hat{n} \rangle_{+} \big) + \big(1 - p' - P_{\mathrm{E}}(t) \big) \delta t\langle \hat{n} \rangle_{-}$
			\State $P_{\mathrm{I}}(t + \delta t) \leftarrow 1 - P_{\mathrm{C}}(t + \delta t) - P_{\mathrm{I}}(t + \delta t)$
			\State $t \leftarrow t + \delta t$
			\EndWhile \Comment{End of second temporal mode}
		\end{algorithmic}
	\end{algorithm}
\end{figure}

Algorithm 1 shows the pseudo-code for the first simulation method where we directly obtain the probability of correct detection $P_{\mathrm{C}}(t)$, error probability $P_{\mathrm{E}}(t)$, and inconclusive probability $P_{\mathrm{I}}(t)$ as a function of time elapsed over the course of the measurement. This method is based on evolving these probabilities in time using Eq. (9) and Eq. (25) from Ref. \cite{nakahira12}. Lines 1 through 10 allow for obtaining the magnitude of the experimental displacement waveform including experimental imperfections. Line 8 enforces the maximum displacement power $R$ as discussed above and Line 9-10 enforces the finite precision (8-bits for our implementation) when performing digital-to-analog conversion from the FPGA. Lines 14-21 evolve $P_{\mathrm{C}}(t)$, $P_{\mathrm{E}}(t)$, and $P_{\mathrm{I}}(t)$ according to Eq. (9) of Ref. \cite{nakahira12} to simulate the first temporal mode of the measurement. Lines 23-31 determine the value of the probabilities at the start of the second temporal mode based on $|\alpha|^{2}$, $P_{\mathrm{I}}$, and $p$ (See Ref. \cite{nakahira12} for details). Lines 33-40 evolve the probabilities through the second temporal mode to give the final values of the error, correct, and inconclusive probabilities.

Algorithm 2 shows the pseudo-code for the second simulation method based on Monte-Carlo sampling. Instead of the probabilities, this method simulates the measurement directly, i.e. as it occurs in the lab based on random sampling from the photon counting distribution. Lines 1-5 fix the particular optimal inconclusive measurement as well as the switching time and value of $v$. Lines 7-50 repeats the simulation of a single measurement (indexed by $i$) $M_{\mathrm{MC}}$ number of times. During a single measurement, Lines 8-12 randomly choose the sign of the true input state $s_{i}=\pm1$, corresponding to $|\pm\alpha \rangle$. Lines 14-17 obtain the displacement magnitude waveform including experimental imperfections. Lines 20-28 simulate the first temporal mode. Line 21 determines the displacement amplitude based on the current total number of photons detected $N$. Line 22 calculates the mean photon number $\langle \hat{n} \rangle$ of the displaced input state detected by the single photon detector including experimental imperfections. Line 23 samples a single detection from a Poisson distribution with mean $\langle \hat{n} \rangle \delta t$ where $\delta t = 1/1024$ is the duration of a single time step (i.e. we discretize the measurement into 1024 discrete time steps). Line 24 enforces on/off detection since for each time step the detection result can only be either vacuum or not-vacuum. Line 27 updates the provisional hypothesis for the input state conditioned on the total number of photons detected thus far. The first temporal mode continues until the time $t_{1}$ is reached. Lines 30-34 determine the displacement phase for the start of the second temporal mode.

\begin{figure}[H]
	\begin{algorithm}[H]
		\caption{Optimal Inconclusive Simulation: Monte-Carlo}
		\begin{algorithmic}[1]
			\scriptsize
			\State Fix: $|\alpha|^{2}$, $P_{\mathrm{I}}$, and $p$ \Comment{Energy, inconclusive probability, and prior probability of more likely input state}
			\State Fix: $\eta$, $\xi$, $\nu$, $R$ \Comment{Experimental Imperfections}
			\State
			\State $t_{1} \leftarrow t_{1}(|\alpha|^{2}, P_{\mathrm{I}}, p)$ \Comment{Eq. 13 in Ref. \cite{nakahira12}}
			\State $v \leftarrow v(|\alpha|^{2}, P_{\mathrm{I}}, p)$ \Comment{Eq. 18 in Ref. \cite{nakahira12}}
			
			\State Fix: $M_{\mathrm{MC}}$ \Comment{Number of Monte-Carlo samples}
			
			\For{$i \leftarrow 1$ \textbf{to} $M_{\mathrm{MC}}$}
			
			\If{$\mathrm{rand} > p$} \Comment{Randomly choose actual input state}
			\State $s_{i} = -1$,
			\Else
			\State $s_{i} = 1$
			\EndIf

			\State
			\State Get $|u(t)|$ from Eq. (1)
			\State $|u_{\mathrm{exp}}(t)| = \mathrm{min}(\sqrt{R}, |u(t)|)$
			\State Transform $|u_{\mathrm{exp}}(t)|$ to nearest bit-value
			\State Transform $|u_{\mathrm{exp}}(t)|$ back to amplitude
			\State
			\State \textbf{Initial:} $N = 0$, $t = 0$, $\hat{s}_{i}(0) = 1$ \Comment{Assume $p\geq0.5$}
			\While{$0 \leq t < t_{1}$} \Comment{Evolve for first temporal mode}
			\State $\beta \leftarrow (-1)^{N}|u_{\mathrm{exp}}(t)|$
			\State $\langle \hat{n} \rangle \leftarrow  \eta \big(|\alpha|^{2} + |\beta|^{2} - 2\xi \beta|\alpha|s_{i} \big) + \nu $
			\State $d \sim \mathrm{Poisson}(n | \langle \hat{n} \rangle \delta t)$\Comment{Vacuum/photon detection}
			\State If $d>1$, set $d=1$
			\State $N \leftarrow N + d$ \Comment{Total number of detected photons}
			\State $t \leftarrow t + \delta t$
			\State $\hat{s}_{i}(t) = (-1)^N$ \Comment{Current hypothesis for input}
			
			\EndWhile \Comment{End of first temporal mode}
			\State
			
			\If{$v \leq 0.5$}
			\State $N_{0} = 1$
			\ElsIf{$v > 0.5$}
			\State $N_{0} = 0$
			\EndIf
			\State
			\State $N \leftarrow N_{0}$
			\While{$t_{1} \leq t < 1$} \Comment{Evolve for second temporal mode}
			\State $\beta \leftarrow \hat{s}_{i}(t_{1})\times(-1)^{N}|u_{\mathrm{exp}}(t)|$
			\State $\langle \hat{n} \rangle \leftarrow  \eta \big(|\alpha|^{2} + |\beta|^{2} - 2\xi \beta|\alpha|s_{i} \big) + \nu $
			\State $d \sim \mathrm{Poisson}(n | \langle \hat{n} \rangle \delta t)$\Comment{Vacuum/photon detection}
			\State If $d>1$, set $d=1$
			\State $N \leftarrow N + d$ \Comment{Total number of detected photons}			
			\State $t \leftarrow t + \delta t$
			
			\If{$\mathrm{mod}(N, 2) = 0$}
			\State $\hat{s}_{i}(t) = \hat{s}_{i}(t_{1})$ \Comment{Hypothesis: $|\pm\alpha\rangle$}
			\Else
			\State $\hat{s}_{i}(t) = 0$ \Comment{Hypothesis: Inconclusive}
			\EndIf
			
			\EndWhile \Comment{End of second temporal mode}
			\EndFor
			\State $P_{\mathrm{C}}(0) \leftarrow p$, $P_{\mathrm{I}}(0) \leftarrow 0$, $P_{\mathrm{E}}(0) \leftarrow 1 - p$
			\State $t \leftarrow 0 $ \Comment{Reset time}
			\While{$0 \leq t < 1$} \Comment{Calculate sampled probabilities}
			\State $t \leftarrow t + \delta t$
			\State $P_{\mathrm{C}}(t) = \sum_{i}\big(\hat{s}_{i}(t)==s_{i} \big) / M_{\mathrm{MC}} $ \Comment{Correct}
			\State $P_{\mathrm{I}}(t) = \sum_{i}\big( \hat{s}_{i}(t)==0 \big) / M_{\mathrm{MC}}$ \Comment{Inconclusive}
			\State $P_{\mathrm{E}}(t) = 1 - P_{\mathrm{C}}(t) - P_{\mathrm{I}}(t)$ \Comment{Error}
			\EndWhile
		\end{algorithmic}
	\end{algorithm}
\end{figure}

Lines 37-50 simulate the second temporal mode in a similar fashion to the first mode, except now the receiver is attempting to discriminate between the hypothesis after the first temporal mode $\hat{s}_{i}(t_{1})$ and the inconclusive outcome denoted by $\hat{s}_{i}(t) = 0$. Lines 38-41 set the displacement field and sample the photon detection for the time step. Lines 44-48 update the provisional hypothesis $\hat{s}_{i}(t)$, which now switches between the hypothesis after the first mode and the inconclusive outcome. This behavior is in contrast to the first temporal mode where the provisional hypothesis switches between the two possible input states. Lines 53-58 calculate the measured values of the correct, error, and inconclusive probabilities based on the value of the hypothesis $\hat{s}_{i}(t)$ and the true value $s_{i}$ across all $M_{\mathrm{MC}}$ Monte-Carlo samples.


%
%
%
\end{document}